# Highly uniform and efficient, broadband meta-beam-splitter/combiner

Saeed Hemayat, Liyi Hsu, Jeongho Ha, and Abdoulaye Ndao*

*Department of Electrical and Computer Engineering & Photonics Center, Boston University, 8 Saint Mary's Street, Boston, MA 02215, USA*

*andao@bu.edu

**Abstract:** Subwavelength planar structured interfaces, also known as metasurfaces, are ultra-thin optical elements modulating the amplitude, phase, and polarization of incident light using nanostructures called meta-atoms. The optical properties of such metasurfaces can be controlled across wavelengths by selecting geometries and materials of the meta-atoms. Given recent technological developments in optical device miniaturization, components for beam splitting and beam combining are sought for use within these devices as two quintessential components of every optical setup. However, realizing such devices using metasurfaces typically leads to poor uniformity of diffraction orders and narrow-band operation. Using a modified version of particle swarm optimization, we propose and numerically demonstrate a broadband, reciprocal metasurface beam combiner/splitter with uniformity>97% and diffraction efficiency>90% in the continuous band from $\lambda$=1525 nm to $\lambda$=1575 nm. The proposed approach significantly extends the current state of the art of metasurfaces design in terms of uniformity, bandwidth, and efficiency, and opens the door for devices requiring high power or near-unit uniformity.

1. **Introduction**

Since the first demonstration of laser by Maiman [1], the demand for high-power lasers has increased remarkably; such lasers are of great importance in industrial applications such as material processing [2], observation and analysis of ultrafast electronic processes [3,4], X-Ray lasers [5], surface hardening [6], quantum electrodynamics [7], and white light generation [8]. Unfortunately, conventional single-channel high-power emitters suffer from various physical limitations like nonlinear and thermo-optical effects and to mode instabilities [9–13]. An alternate approach is using beam combiners [10,14–17] to combine multiple lasers in an array-like fashion to obtain a high-power beam, avoiding the difficulties above associated with single emitters.

In their simplest form, beam combiners, can be considered beam splitters used in reverse with the phase and amplitude of the beams adjusted [10]. This can also be explained from the antenna's perspective through the reciprocity theorem, as an antenna (and also an array of antennas) transmits and receives with the same radiation pattern (antenna gain is the same whether it operates in transmitter or receiver mode) [18]. As a result, if a beam splitter generates highly uniform diffraction orders in the farfield, one can expect the same device to act as a beam combiner with high diffraction efficiency if used with reverse-propagating fields. Dammann gratings, which are binary phase holograms, have been used over decades as a trustworthy candidate for generating uniform diffraction orders in the far-field [19,20] and have been utilized to realize beam splitter/combiners, multi-imaging systems, laser beam shaping, and laser parallel micro-processing [21–29]. Conventionally, Dammann gratings were fabricated by controlling the etch depths to vary the phase, but this approach gives rise to fabrication difficulties as it requires multiple lithographic procedures [30]. Therefore, an alternative method is highly desirable for implementing Dammann gratings to achieve near-perfect uniformity and diffraction efficiencies in beam splitters/combiners.

As current technology rapidly advances toward the complete miniaturization of optical components, a fully integrable beam splitter/combiner is highly demanding. Metasurfaces emerge as an excellent platform to achieve the miniaturization of optical components due to their localized phase control. In recent years both plasmonic and dielectric metasurfaces have been used to achieve various optical functionalities such as metalenses [31–34], Bessel beam generation and super-resolution focusing [35,36], power-limiters [37], large field of view structured light projection [38], carpet cloaking [39,40], holograms [41], sensing [42–46], anisotropic metasurfaces [47–50], and beam splitters and combiners [51–56]. However, because of the phase discretization on the metasurfaces, which gives rise to unwanted high order components (diffraction orders in the far-field), most existing metasurface-based beam splitters/combiners suffer from low uniformity, particularly in the form of stronger zeroth-order diffraction with respect to other diffraction orders, or low diffraction efficiency, which highly affects the diffraction orders in the far-field [53–55,57–61]. On the other hand, it is challenging to correct this phase discretization in a theoretical framework due to a large number of variables (rotation of each meta-atoms) in geometrical metasurfaces.

Here we present a broadband, metasurface beam splitter/combiner with diffraction efficiencies of 93% (2D), uniformity of 98.6% as a beam splitter, as well as diffraction efficiencies 88.6% as a beam combiner at the central wavelength of $\lambda$=1550 nm. Our design is based on a Dammann grating with the primary goal of obtaining highly uniform diffraction orders in the far-field. In addition to the commonly adapted geometric phase, we further propose and adapt a modified version of particle swarm optimization (PSO) as a global optimization strategy to compensate for the phase discretization effects which in turn gives rise to undesired higher diffraction orders. Our adaption jointly maximizes both diffraction efficiency and uniformity of diffraction orders, and successfully finds an optimum solution in huge parameter space (729 meta-atoms with rotation angles between 0 and 180 degrees). Numerical results show a 16.5% and 15.5% increase in diffraction efficiency and uniformity of the 2D beam splitter and a 12.2% increase in the diffraction efficiency of the 2D beam combiner following the optimization stage. Our approach avoids the complex structures resulting from counterpart methods such as inverse design because it is based on the optimization of rotations of meta-atoms, thus lowering the fabrication complexity. This generic method is highly scalable and can be easily applied to achieve a higher number of uniform diffraction orders in the far-field. Furthermore, the proposed device is broadband with uniformity>97% and diffraction efficiency>90% (diffraction efficiency>87% for the beam combiner) over the continuous band from $\lambda=1525$ nm to $\lambda=1575$ nm. The proposed design and optimization procedure paves the way for miniaturized beam splitting and combining, making it a potential centerpiece of many applications which require high uniformity, like, quantum-photonics, depth sensing and facial recognition, or applications that require high power, avoiding the nonlinear effects and mode instabilities in conventional high-power lasers.

## 2. Structure design and operation principles

As previously elucidated, a beam combiner can be designed by optimizing its beam-splitting capabilities. Figure 1 shows an exemplary 3×3 beam splitting/combining application using our proposed metasurface. As the incident beam goes toward the metasurface along the +z direction, as shown in Fig. 1(a), the metasurface splits the beam into 3×3 diffraction orders. Conversely, if nine laser beams are incident at the exact diffraction angles but in the reverse direction, as depicted in Fig. 1(b), the metasurface will combine the nine beams into one high-power beam.

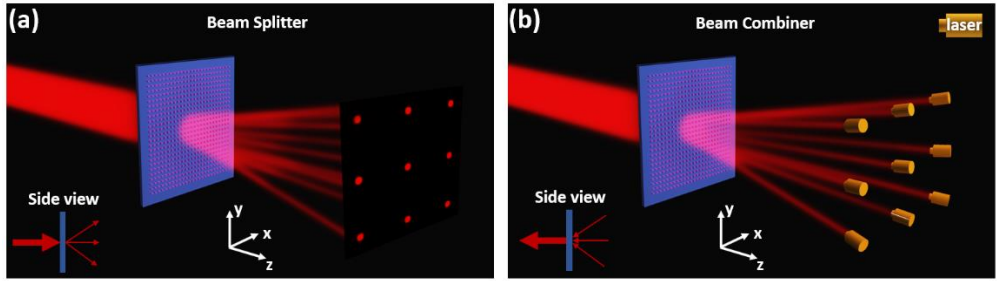

Fig. 1. Concept figure showing the operation principle of the reciprocal beam splitter/combiner. (a) beam splitter; (b) beam combiner.

Given the binarized nature of Dammann gratings, as proposed in [19], the only free parameter in the design of Dammann gratings are transition points where the phase changes from 0 to π or vice-versa. Here, we aim to design a 1D (1×7) and a 2D (3×3) beam splitter, and consequently, there exist three and one transition points ($N_d=2N+1$, where $N_d$ is the total number of desired diffraction orders, and $N$ is the number of transition points) for the 1×7 and the 3×3 case, respectively. The diffraction orders in the far-field can be obtained by taking the Fourier transform of the device's phase profile. As a result, one can write the diffraction orders obtained using Dammann grating as the following [19,55]:

$$A_0 = 2\sum_{i=1}^{N}(-1)^i(\xi_i - \xi_{i+1}), \tag{1}$$

$$A_n = \frac{1}{n\pi}\sum_{i=1}^{N}(-1)^i(\sin 2n\pi\xi_i - \sin 2n\pi\xi_{i+1}). \tag{2}$$

Here $A_0$ and $A_n$ are amplitudes of the zeroth-order and $n^{th}$-order diffraction spots, respectively, and $\xi_i$ are the transition points. To achieve uniform 1×7 ($N=3$) diffraction orders in the far-field, the three transition points along the x-axis in $0 \leq x \leq 1/2$ range are $\xi_1 = 0.2405$, $\xi_2 = 0.3655$, and $\xi_3 = 0.4380$ as shown in Fig. 2(a). To construct a 2D theoretical phase on the metasurface capable of generating uniform 3×3 diffraction orders, first, we design the phase vector on the metasurface to obtain uniform 1×3 diffraction orders in the far-field, where the only transition point while $0 \leq x \leq 1/2$ is $\xi_1 = 0.18905$. Consequently, the 2-dimensional theoretical phase to achieve 3×3 diffraction orders can be obtained by multiplying the 1D phase vector (for 1×3 splitter) by its transpose vector, which will result in a phase profile shown in Fig. 2(b).

To achieve such a binarized phase profile with a metasurface, we use Pancharatnam-Berry (P.B.) phase to design the proposed geometrical metasurface; the phase shifts between adjacent meta-atoms are induced via rotation of meta-atoms [35,62,63]. None of the other geometrical parameters (e.g., height, $R_1$, $R_2$, or period as shown in Fig. 2(c)) change throughout the entire design procedure. This feature prohibits generating entirely random structures that are challenging to fabricate, which occurs in most of the structures generated by inverse-design. To perform the numerical simulation, we used FDTD module of ANSYS LUMERICAL for both unit cell and full-wave simulations of the whole structure throughout the optimization procedure. For optimization procedures, LUMERICAL was linked to MATLAB by introducing MATLAB API to LUMERICAL, which enabled us to gain full control of LUMERICAL from inside MATLAB. In each iteration, a set of rotation angles are generated for each of 729 meta-atoms, evaluating the cost function.

To find the optimum values of the fixed geometrical parameters for the unit cell, a comprehensive sweep of different geometrical parameters is done using the Finite-difference Time-domain (FDTD) method to achieve the maximum transmission and the maximum conversion efficiency between the incident circularly polarized (C.P.) beam and the orthogonal output C.P. beam. The metasurface is designed for the center wavelength of $\lambda$=1550 nm using amorphous silicon (α-Si). Each unit cell comprises an α-Si elliptical cylinder on a $SiO_2$ substrate (Optical constants of silicon and $SiO_2$ are extracted from Palik [64]). Since each unit cell is a Pancharatnam-Berry Optical Element (PBOE), in the ideal case, PBOEs should act as a half-waveplate and transform the incident LCP/RCP beam to its orthogonal output polarization [36,65,66]. The sweep of different geometrical parameters, the maximum transmission and conversion efficiency is achieved for a unit cell with $P$=750 nm, $R_1$=215 nm, $R_2$=140 nm, and $H$=850 nm. The unit cell of the metasurface and the corresponding electric field is shown in Fig. 2(c) and Fig. 2(d), respectively. Intuitively a smaller period gives rise to more accurate results, as the number of sampling points in the phase space is increased and hence leads to a less discretized phase response; However, to mitigate the fabrication complexities, the minimum distance between two adjacent meta-atoms must be chosen carefully (typically more than 50 nm). The analytical equation for the transmission of a rotated unit cell can be derived using $T(\theta)=R(-\theta)T_1R(\theta)$, where $R(\theta)$ is the rotation matrix [63]:

$$T(\theta) = \begin{pmatrix} \cos\theta & -\sin\theta \\ \sin\theta & \cos\theta \end{pmatrix} \begin{pmatrix} T_o & 0 \\ 0 & T_e \end{pmatrix} \begin{pmatrix} \cos\theta & \sin\theta \\ -\sin\theta & \cos\theta \end{pmatrix}. \tag{3}$$

Here, $T_o$ and $T_e$ represent the respective complex transmission coefficients when the polarization of incoming light is aligned toward the principal axes of the meta-atom, and $\theta$ is the rotation angle. Eventually, if the incident light is circularly polarized, the expression for the transmitted electric field is given as [35,63]:

$$E_{L/R}^t = \frac{T_o+T_e}{2}\cdot\frac{1}{\sqrt{2}}(\hat{e}_x \pm i\hat{e}_y) + \frac{T_o-T_e}{2}\cdot\frac{1}{\sqrt{2}}e^{i(2\theta)}(\hat{e}_x \pm i\hat{e}_y). \tag{4}$$

where $\hat{e}_x$ and $\hat{e}_y$ are electric field components along $x$ and $y$ directions. The second term in the equation above shows that one can induce a phase equal to two times the rotation of each element by rotating each unit cell while using the C.P. light. Rotation of each meta-atoms can be obtained by dividing the desired phase on the metasurface by two, i.e., $\theta=\varphi(x,y)/2$ where $\varphi$ is the phase profile of the metasurface and $\theta$ is the rotation of each meta-atoms. The Dammann metasurface beam splitter capable of generating 3×3 uniform diffraction orders in the far-field is shown in Fig. 2(b).

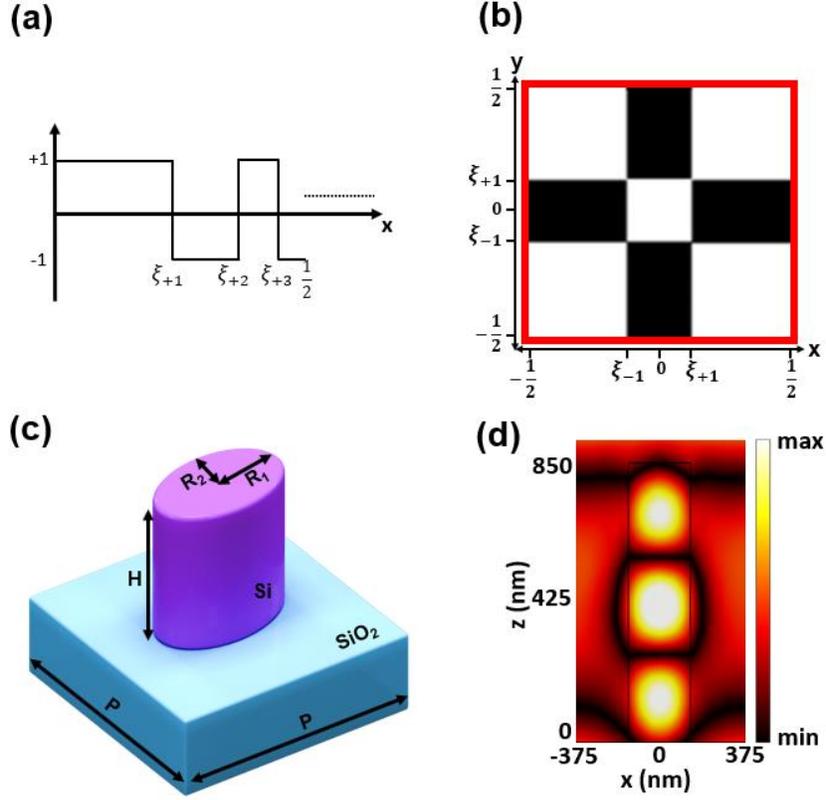

Fig. 2. (a) Phase profile of a unit cell of 1×7 Dammann beam splitter spanned from 0 to +1/2, 1 and -1 on the vertical axis correspond to the relative phases of π and 0, (b) phase profile of a unit cell of 3×3 Dammann beam splitter with white and black regions corresponding to π and 0 phase, respectively, (c) a unit cell of the designed metasurface, (d) electric field intensity in one unit cell at λ=1550 nm.

To quantitatively assess the performance of the device, two criteria have been selected: diffraction efficiency, given by total power concentrated in the desired diffraction orders, and uniformity $U$, defined as:

$$U = 1 - \frac{\max(g) - \min(g)}{\max(g) + \min(g)}, \qquad (5)$$

where vector $g = [g_1, g_2, g_3, g_4, g_5, g_6, g_7, g_8, g_9]$, in which each $g_{1-9}$ represents the fraction of power concentrated in (-1,-1), (-1,0), (-1,+1), (0,-1), (0,0), (0,+1), (+1,-1), (+1,0), and (+1,+1) diffraction orders, respectively. As a result, uniformity can be obtained by finding the minimum and maximum of vector $g$ and using Eq. (5). The total diffraction efficiency of the metasurface, defined as the sum of all the power concentrated in the desired orders is defined as $G$. The results obtained directly from theoretical formulations are presented in Fig. 3. The diffraction efficiency for the beam combiner is defined as the fraction of total power concentrated in the zeroth-order diffraction order. The diffraction efficiency ($G$) of the 1D beam splitter and beam combiner is 78.5% (and 77.9% uniformity) and 81.2%, respectively. For the 2D designs, the diffraction efficiencies for the beam splitter and beam combiner are respectively 77% (and 83% uniformity) and 76.4%. The low values of uniformity and diffraction efficiency for the theoretical metasurface design are attributed to discretization effects and sidelobes, which arise due to the interference between the radiation patterns of each array element (meta-atoms). Consequently, the results obtained from the theory of Dammann gratings are far from ideal if realized using metasurfaces due to the phase discretization in the regions where there should be no transition points, and hence no phase changes are allowed.

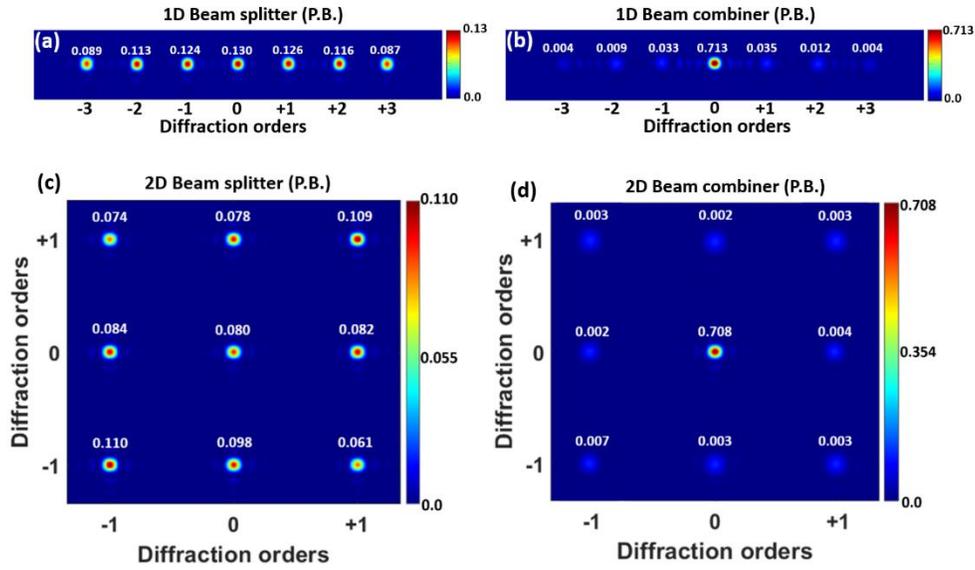

Fig. 3. Diffraction orders obtained from theoretical designs (following P.B. phase and before optimization) at λ=1550 nm. (a) The 1D beam splitter with 77.9% uniformity and 78.5% diffraction efficiency, (b) 1D beam combiner with 81.2% diffraction efficiency, (c) 2D beam splitter uniformity=83% and 77% diffraction efficiency, and (d) 2D beam combiner with 76.4% diffraction efficiency.

## 3. Optimization procedure

One of the most successful categories of global optimization is evolutionary algorithms [67], such as genetic algorithms [68], PSO [69], ant colony [70], artificial bee colony [71], and many more that have been used with great success in a wide variety of applications like image processing and enhancement [72–74], industry [75], and medical applications [76]. The significant advantage of PSO over other evolutionary algorithms, such as genetic algorithm, besides its ease of implementation and faster convergence speed, lies in the memory of the particles and the information flow between them. Each particle (agent) in PSO has memory and shares information with all other particles while obeying some general rules (self-organization). However, as the size of the current problem is extremely large (729 meta-atoms, each accepting a rotation value between 0 and 180), one must carefully modify the conventional PSO to prevent premature convergence. PSO can be summarized in three steps: 1. slight movement of each particle along the same direction as its previous velocity vector, 2. slight movement of each particle toward its own best cost, and 3. slight movement of each particle toward the global best of all particles. The result would be a vector summation of all the vectors obtained from the three steps mentioned above. These steps can be mathematically written as the following:

$$v_i(t+1) = wv_i(t) + r_1c_1[p_i(t) - x_i(t)] + r_2c_2[g_i(t) - x_i(t)], \tag{6}$$

where, $v_i(t+1)$ is the new velocity vector of the $i^{th}$ particle, $w$ is the inertia coefficient (typically less than one and generally takes values from 0.4 to 0.9) which determines how much the particle tends to hold its current movement direction, $v_i(t)$ is the current velocity vector of the particle, $r_1$ and $r_2$ are random numbers, $c_1$ and $c_2$ are personal and global learning coefficients, respectively. The first term on the right-hand side of Eq. (6) represents the tendency of the particle to move along the same direction as its current velocity vector. The second term determines the movement along the best personal experience of the present particle, and the third term illustrates the movement along the global best experience of all particles. The new position of the particle can be obtained as follows:

$$x_i(t+1) = x_i(t) + v_i(t+1), \tag{7}$$

where $x_i(t+1)$ is the new position of the particle. To ensure the best balance between exploration and exploitation capabilities of the algorithm, constriction coefficients are used for the inertia, personal, and global learning coefficients [77]. However, as discussed above, due to the large number of variables that can take value in an uncountable set, the optimization algorithm is highly likely to get stuck in local minima.

To prevent this, we have modified the conventional PSO by adding a layer of conservative local search. This is implemented by defining a saturation threshold for the number of function evaluations. We define saturation as the cases in which the cost function does not change after a certain number of function evaluations. First, the conventional PSO runs, and if it reaches the saturation threshold, instead of terminating, random noises within a specific range are added to both positions and velocities of the particles, and then again, the algorithm enters a new PSO procedure, this time with the updated noisy positions and velocities. This noise distribution is generated using Latin hypercube sampling (LHS) of a multivariate uniform distribution [78] to ensure that the noise is distributed uniformly all over the local search space, where a cumulative density function (CDF) is divided into non-overlapping regions. One value is randomly chosen from each region, which inherently ensures that at least one value is sampled from every region (as opposed to simple random sampling) [79].

The noise injection phase compensates for the fact that in conventional PSO, the velocities gradually decrease in each run to help with the exploitation. However, this increases the probability of premature convergence, i.e., if the initial population was not covering the whole space and the particles were gathered in clusters, they would converge toward a local minimum. Due to the degraded velocity vector, the particles will never escape that local minimum. As a result, the algorithm shows no exploration capabilities. The range for these random noises must be chosen carefully, as we know that the optimum solution lies in the neighboring theoretical design. We started the optimization procedure to optimize the results obtained from the theory. Consequently, there is no need to search regions far from the current global best achieved with the conventional PSO occurring before noise injection. It is worth noting that the position and velocity of particles lie in different spaces, and as a result, the noise and its range are different for positions and velocities. After the noise injection phase, 1% of all particles are selected randomly in each saturation threshold step. Their position will be mirrored spatially to help with the local exploration (apart from the principal coordinate axis of the problem, new local coordinate axes are defined in each saturation threshold step and their centers are set at the center of each particle cluster (local minima) just before the noise injection phase). As a result of increased local exploration and exploitation capabilities of the algorithm, they act synergistically to get out of a local minimum. To maximize the sum of desired diffraction orders and minimize the difference between diffraction orders, the cost function of the problem is defined below:

$$Cost\ Function = \sum_{k=1}^{n^2} \frac{1}{g_k} + \left[ \sum_{i=1}^{n^2} \sum_{j=1}^{n^2} \frac{|g_i - g_j|}{(g_i + g_j)^2} \right], \tag{8}$$

where, $g_{i,j,k}$ are diffraction orders. The pseudocode of the modified PSO is presented below:

**ALGORITHM 1. PSEUDOCODE FOR THE PROPOSED MODIFIED PSO**

```
Begin procedure;
    Pseudorandom generation of particles;
    for (particle i∈N):
        Calculate Cost_function(i);
        Initialize inertia_coeff, constriction_coeffs, rand_vel_noise, rand_pos_noise, and
        random_noise_coeff;
        for each particle:
            pBest(i) ← particle(i).BestPosition;
            if (Cost_function(i)<pBest):
                pBest(i) ← Cost_function(i);
            end
        end
        gBest ← best cost of all particles;
        for each particle:
            Calculate particle(i).velocity using Eq. (6);
            Update the particle(i).position using Eq. (7);
        end
        Update inertia_coeff;
        if (saturation_threshold=true): /*Saturation threshold layer*/
            for (particle j∈N)
                rand_pos_noise ← min(particle.pos(j)) + [(randperm(num_particles)'-…
                rand(:,j))/(num_particles.* (max(particle.pos(j))-min(particle.pos(j)))];
                /*rand() generates random dist*/
                /*randperm() generates random perm. of integers*/
                rand_vel_noise ← min(particle.vel(j)) + [(randperm(num_particles)'-…
                rand(:,j))/(num_particles.* (max(particle.vel(j))-min(particle.vel(j)))];
            end
            rand_pos_noise ← rand_pos_noise* rand_noise_coeff;
            rand_vel_noise ← rand_vel_noise* rand_noise_coeff;
            add rand_pos_noise to particle(i).pos;
            add rand_vel_noise to particle(i).vel;
            local_center ←Find average x and y coordinates of particle clusters;
            selected_mirror_particles ←Select 1% of each cluster;
            for each selected_mirror_particles
                selected_mirror_particles.x← -(selected_mirror_particles.x);
                selected_mirror_particles.y← -(selected_mirror_particles.y);
            end
            Update rand_noise_coeff;
            Run PSO's main loop; /*with updated positions and velocities*/
        end
        Check the termination criteria; /*if gBest is less than termination threshold*/
    end
END
```

*gBest* in the above pseudocode is a global property, and no index has been used to define it since it is not assigned to a particular particle. The normalized cost function for the 1D and 2D beam splitter/combiner is plotted with respect to the number of function evaluations. As shown in Fig. 4(a), the algorithm starts to converge for the 1D design after approximately 4500 function evaluations, and the operation is terminated early (NFE<10000 for 1D case); as uniformity reaches 100%. For the 2D case, however, the convergence is achieved after approximately 14000 function evaluations, which is expected due to the larger problem size and more significant local minima. In the 2D case, however, the algorithm faces multiple local minima, and the saturation-threshold layer runs numerous times.

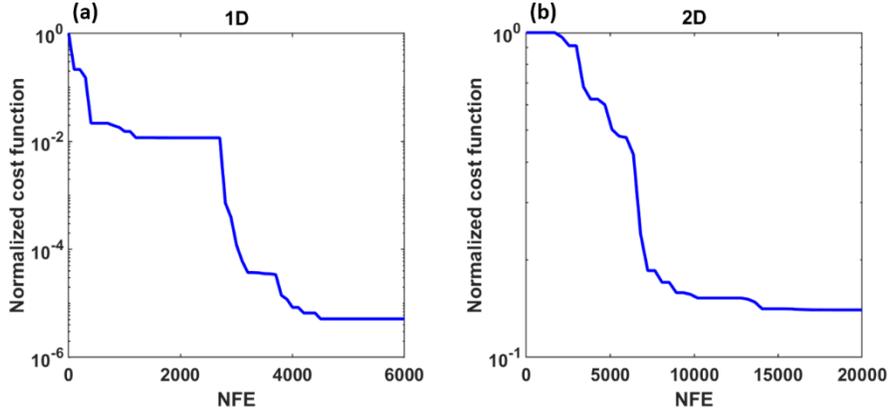

Fig. 4. Cost function values for 1D and 2D metasurface Dammann gratings, plotted with respect to the number of function evaluations. (a) 1D and (b) 2D.

Since the current design is reciprocal, the optimization procedure must be done only once for one of the cases (either for the beam splitter or the beam combiner). This is because the device can be viewed as a 2D phased array of antenna elements (meta-atoms) arranged adjacent to each other with different phases. Therefore, according to the reciprocity theorem, the antenna gain is the same in transmission and receive modes [18].

### 4. Optimized metasurface beam splitter/combiner

This section presents the results obtained for the optimized 1D and 2D beam splitter/combiner (Fig. (5)) at $\lambda=1550$ nm. For the optimized design, one can observe the improvement in diffraction efficiency and uniformity of the diffraction orders as compared to the designs obtained directly from theory. The diffraction orders for the optimized 1D beam splitter are perfectly uniform, and 100% uniformity is achieved (22.1% increase compared to the unoptimized structure), as shown in Fig. 5(a). Also, 96.6% of the total diffracted power is concentrated in these seven diffraction orders (diffraction efficiency=96.6%). In the case of the 1D beam combiner (shown in Fig. 5(b)), 92% diffraction efficiency is achieved, which is 10.8% higher than the unoptimized design. For the 2D beam splitter, 98.5% uniformity and 93.5% diffraction efficiency are reached (Fig. 5(c) and (e)), a 15.5% increase in uniformity and 16.5% increase in diffraction efficiency with respect to the unoptimized design, while an 88.6% diffraction efficiency corresponds to a 12.2% improvement over the unoptimized case for the beam combiner (Fig. 5(d) and (f)). Both the theoretical and optimized results are summarized in Table 1. It is noteworthy that the sum of all diffraction orders equals one. However, the sum is less than one in Fig. (3) and Fig. (5), as in these figures only the desired orders are shown. The orders in the extended region are not shown (i.e., if we extend the whole area in Fig. 3(c) and (d) or Fig. 5(c-f) to ±3 diffraction orders, then very dim diffraction orders will appear to be visible where the remaining power is concentrated in these orders. The same argument also applies to the 1D cases).

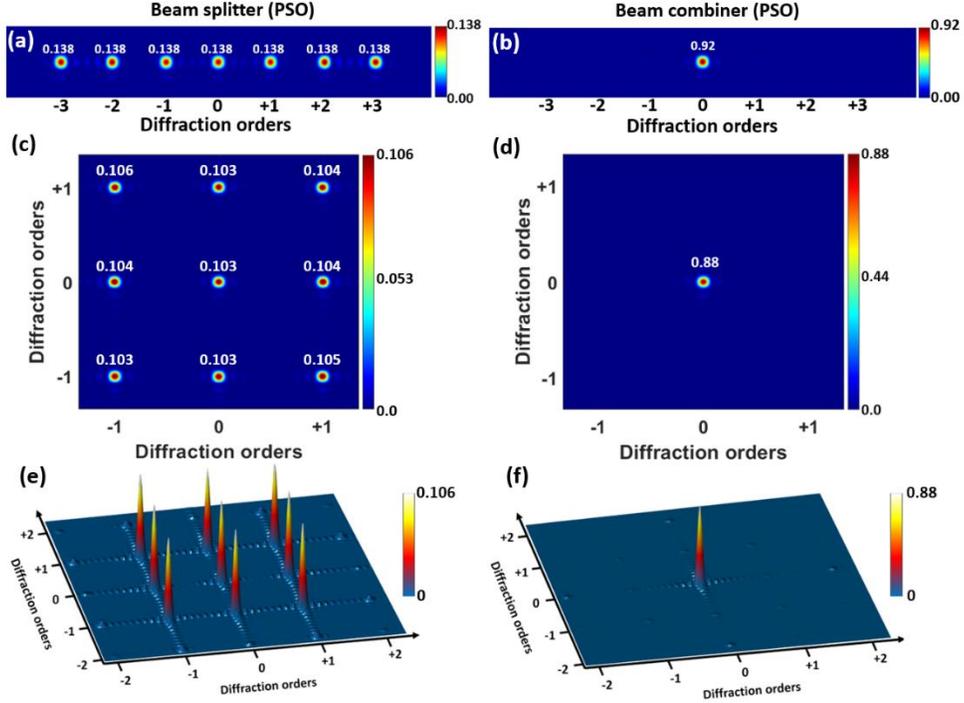

Fig. 5. Diffraction orders at λ=1550 nm after optimization. (a) 1D beam splitter with 100% uniformity and 96.6% diffraction efficiency, (b) 1D beam combiner with 92% diffraction efficiency, (c) 2D beam splitter with 98.5% uniformity and 93.5% diffraction efficiency, (d) 2D beam combiner with 88.6% diffraction efficiency. For better clarification of the concept, a 3D representation of the diffraction orders for the 2D beam splitter and beam combiner are shown in (e) and (f), respectively.

Table 1. Diffraction efficiency and uniformity of diffraction orders in the far-field for 1D and 2D beam splitter/combiner.

| Name | Diffraction efficiency | Uniformity |
|---|---|---|
| 1D BS, Theory | 78.5% | 77.9% |
| **1D BS, Optimized** | **96.6%** | **100%** |
| 1D BC, Theory | 81.2% | - |
| **1D BC, Optimized** | **92%** | - |
| 2D BS, Theory | 77% | 83% |
| **2D BS, Optimized** | **93.5%** | **98.5%** |
| 2D BC, Theory | 76.4% | - |
| **2D BC, Optimized** | **88.6%** | - |

To analyze the broadband characteristics of the optimized design, we define the operating bandwidth as the spectral region where diffraction efficiency is greater than 90% and uniformity is greater than 97%. The design is considered broadband only if it meets these two criteria simultaneously. The proposed broadband design characteristics are illustrated in Fig. (6). Even though the broadband requirements are chosen conservatively, the proposed design satisfies both criteria over a 50 nm bandwidth. A consistent response has been achieved over this bandwidth range, except for slight variations in the intensity of diffraction orders at each wavelength and a slight variation in the angles of the diffraction orders. It is worth noting that if we decrease the broadband limits to 85% and 92% for diffraction efficiency and uniformity, a bandwidth of almost 100 nm is also achievable. This broadband behavior is attributed to the metasurface's design based on dielectric waveguides. The broadband characteristics of the proposed design are summarized in Table 2.

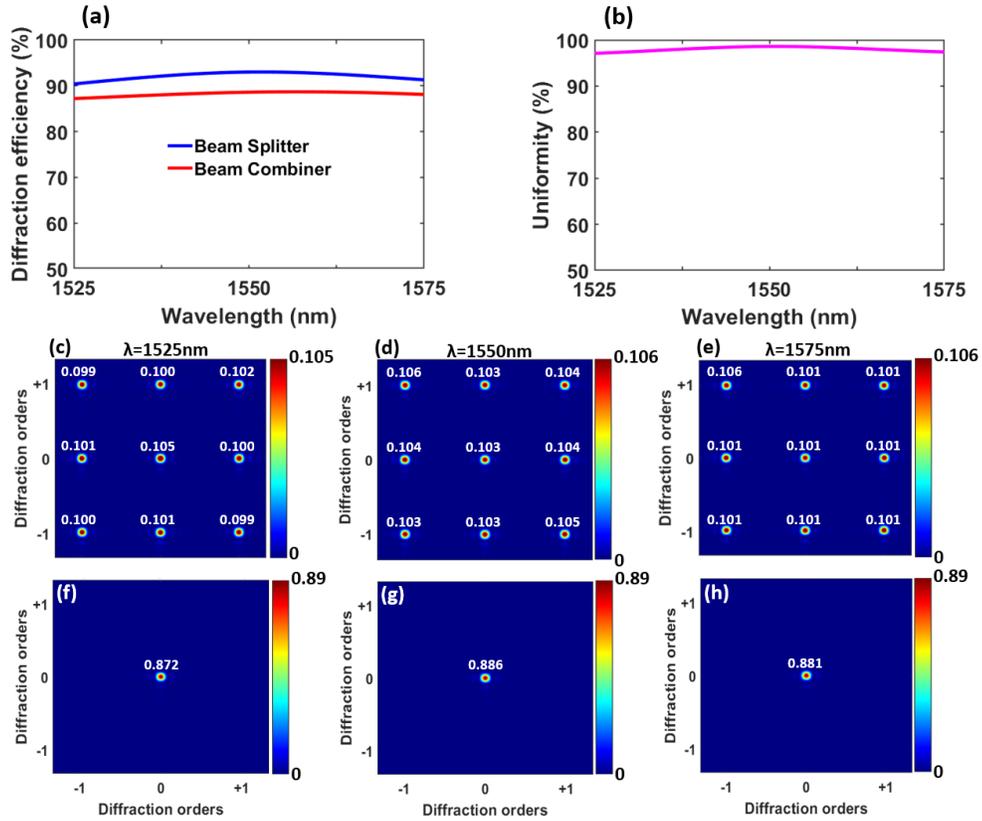

Fig. 6. Broadband characteristics of the proposed beam splitter/combiner. (a) Diffraction efficiency and (b) uniformity of the diffraction orders with respect to changing the wavelength. (c)-(e) diffraction orders at $\lambda$=1525 nm, $\lambda$=1550 nm, and $\lambda$=1575 nm, respectively, while operating as a beam splitter, (f)-(h) diffraction orders at $\lambda$=1525 nm, $\lambda$=1550 nm, and $\lambda$=1575 nm, respectively, while operating as a beam combiner.

Table 2. Broadband characteristics of the proposed 2D beam splitter/combiner

| Name | λ=1525 nm | λ=1550 nm | λ=1575 nm |
|---|---|---|---|
| 2D BS, Diffraction efficiency | 90.4% | 93.5% | 91.4% |
| 2D BS, Uniformity | 97.1% | 98.5% | 97.5% |
| 2D BC, Diffraction efficiency | 87.2% | 88.6% | 88.1% |

## 5. Conclusions

We proposed and numerically demonstrated a metasurface beam combiner/splitter generating high diffraction efficiency (diffraction efficiency>90% for beam splitter and diffraction efficiency>87% for beam combiner) and near unit uniformity (uniformity= 97%) in the continuous wavelength range from 1525 nm to 1575 nm. Such performances are achieved using a modified version of particle swarm optimization. To the best of our knowledge, this is the highest uniformity, and highest diffraction efficiency reported to date. The proposed approach significantly extends the current state of the art of metasurfaces design in terms of uniformity, bandwidth, and efficiency, and opens the door for devices requiring high power or near unit uniformity.